\documentstyle[epsf]{espart}
\begin{document}
\begin{frontmatter}
\title{ Nuclear Shell Model Calculations
        with Fundamental Nucleon-Nucleon Interactions}

\author[drexel]{Xing-Wang Pan},
\author[stony]{T.T.S. Kuo},
\author[drexel]{Michel Valli\`eres} \\ ~~and~~
\author[drexel]{Da Hsuan Feng}
\address[drexel]{Department of Physics, Drexel University,
Philadelphia, PA 19104, USA}
\address[stony]{Department of Physics,
State University of New York at Stony Brook, \\
Stony Brook, NY 11794, USA}

\begin{abstract}
Some fundamental Nucleon-Nucleon interactions and their applications
to finite nuclei are reviewed.  Results for the few-body systems and
from Shell-Model calculations  are discussed and
compared to point out the advantages and disadvantages of the different
Nucleon-Nucleon interactions.

\end{abstract}
\end{frontmatter}

\section{Introduction}
 A major task in our microscopic study of nuclei is
 to describe nuclear properties with fundamental nucleon-nucleon
 interactions. First among these properties is the
 experimental binding energy (of nuclear few-body
 or many-body systems). The prediction of this property remains a goal for
 any microscopic nuclear structure model; it remains a severe test
 of the underlying dynamics. The Shell-Model, in addition,
 also aims at predicting the spectroscopic
 details unraveled by the experiment.

 The first step towards a fundamental description of the nuclear few-body or
 many-body systems using the nucleon degrees of freedom
 is the establishment of the nucleon-nucleon ($NN$) interaction derived
 from the underlying dynamics; this can take various forms.
 From a fundamental point of view, one uses the derived
 realistic $NN$ potentials to directly solve Faddeev equations for
 the eigenvalues of the nuclear
 few-body systems, or to determine a model-space dependent
 microscopic nuclear interaction for the nuclear
 many-body systems (to be used in the  nuclear Shell-Model or the
 Bruckner Hartree-Fock approach).
 For nuclear many-body systems, starting from a fundamental
 stand point, one can take into account the
 strongly repulsive bare $NN$
 interaction by first selecting an effective model space and
 then constructing an effective Hamiltonian (nuclear $G$ reaction matrix)
 to describe the complicated many-body bound states.

 Effective Hamiltonians can also be defined based on
 simple parameterization of the $NN$ interaction, with parameters
 fitted to the few body systems or directly for the large systems.
  This alternative avoids the use of the bare $NN$ potentials.
 Effective Hamiltonians can also be defined to account for some
 characteristic properties (for instance, pairing correlation and/or
 symmetry issues \cite{elliott}) of a many-body system.

 At times, the effective Hamiltonians are the result of global fitting
 \cite{brown}; this leads to the empirical Shell-Model approach.
  Let us say at the outset that this approach
 has achieved  a very successful unified description
 for most of $s-d$ shell stable nuclei \cite{brown2}.
 Yet, the Shell-Model has yet to produce a
 deep microscopic understanding. Besides,
 we know that the huge unfeasible Shell-Model spaces in
 the traditional large-scale Shell-Model calculations
 make it difficult to perform
 systematic calculations for nuclei beyond those in the
 $f\mbox{-}p~$ shell (except for some heavier nuclei near close shells).
 These nuclei often require huge scale Shell-Model calculations
 involving the best recent algorithms
 to deal with the large model spaces, while so many
 two-body matrix elements (195 independent 2-body matrix elements
 for fp-shell) cannot globally be fitted as easily as in sd-shell.
 Beyond the $f\mbox{-}p~$ shell,
 some dramatic truncation of the Shell-Model is required
 to even render the effective Shell-Model possible (for instance,
 broken-pair Shell-Model \cite{broken}). Therefore,
 the microscopic Shell-Model study with $NN$ interactions
 is not only tackling a fundamental nuclear many-body problem, but
 it may provide helpful information for the empirical approach.

 In this report, we will analyze the physics implied by
 different $NN$ interactions. We will carry out simple
 Shell-Model calculations
 for modest valence-particles system (more specifically,
 $^{18}O$, $^{18}F$ and $^{18}Ne$ ),
 where exact calculations can eliminate some uncertainties,
 starting from various $NN$ interactions (i.e, using the
 $NN$ interactions as input to obtain Bruckner's G-matrix as
 microscopic effective two-body interactions). We argue that different
 $NN$ forces can give substantial discrepancies in nuclear low-lying
 spectroscopy. The aim of our calculations
 is to provide us with some lessons of and insights about the
 microscopic study of the nuclear dynamics.

 This report is organized as follows:
 In sect 2, we will briefly review some modern $NN$ interactions
 and the physics they encompass.
 Sect.3 will review the present status of the few-body and many-body
 calculations based on realistic $NN$ forces.
 In sect. 4, attention will be paid to
 the nuclear many-body systems.
 The process to get effective interactions from $NN$ forces
 is briefly mentioned.
 In sect. 5, we will first mention the recently developed
 {\em Drexel University
 Shell Model} (DUSM) algorithm and code, then describe some
 of the results and follow up by a brief discussion of the Shell-Model
 calculations. Finally,  comments will be given  in sect. 6.

 \section{Nucleon-Nucleon Interaction}
 The description of the
 nucleon-nucleon interaction remains one of the most fundamental
 themes in low-energy hadron physics. The resulting force has
 important consequences for nuclear physics.
 With the advent of quantum chromodynamics (QCD),
 one would naturally hope to derive
 the $NN$ force which is based on the quark-gluon dynamics. Typically,
 under the assumption of spontaneous breaking of
dynamical symmetry and with 't Hooft's
large $N_{c}$ expansion approximation,
and by formally integrating out the quark and gluon fields,
one can obtain an effective chiral lagrangian of low-energy
hadron dynamics, namely, a ``derivation''
of skyrme-like lagrangian in the low-energy regime of QCD \cite{leff}.
Recently, a derivation from the QCD lagrangian to a hadron dynamics,
lagrangian, which contains QCD string structured mesons and baryons,
has been achieved. Based on the derived hadron dynamics lagrangian,
the pseudoscalar coupling constant was theoretically estimated \cite{pan}.
Such ``derivations'' provide a justification
for the effective meson model of nucleon-nucleon interaction.

 Due to the non-perturbative
 nature of low-energy QCD, the description of the
 $NN$ force is carried out in practice directly via
 the nucleon-meson process, i.e., using
 the meson degree of freedom to describe the interaction between
 two free nucleons.
 Using this meson based description, the nucleon-nucleon interactions
 follow from three different approaches:
 second order perturbation theory (e.g., Yukawa potential) \cite{yukawa},
 dispersion theory (e.g. Paris potential) \cite{paris},
 and the field-theoretical meson-exchange model for
 the $NN$ interaction (e.g., Bonn potential \cite{bonn} and Nijmegen
 potential \cite{Nijmegen}).
 There are also earlier and more  phenomenological approaches
 to the nuclear potential; in particular we quote the
 hard-core Hamada-Johnston (HJ) potential \cite{hamada},
 the Reid soft-core (RSC) Potential \cite{reid} and the
 Argonne $v_{14}$ \cite{v14}.  We will come back to these later.
 We mention that
 chiral symmetry perturbation theory has recently been used to explore certain
 specific terms in the  $NN$
 interactions (for example, the three-nucleon force)
 \cite{weinberg}.

 The $NN$ force can be effectively divided into three interaction ranges:
 long-range one-pion-exchange
 part ($r\stackrel{>}{\sim} 2 fm$),
 intermediate-range attractive part ($1\sim2 fm$) and the short-range
 repulsive part ($r \stackrel{<}{\sim}1 fm $).
 This division follows naturally from the physics involved at
  these various distances.
 The long-range part presents fewer ambiguities since it
 is dominanted by the
 one-pion exchange (OPE) effect. In fact, various realistic
 potential models have very similar long range part;
 they differ however at shorter distances, and in particular
 in their intermediate-range part, which
 are supposedly due to the contribution from heavier mesons (or pion
 resonance) or two-pion exchanges (TPE).
 For instance, the Paris potential determines the TPE from $\pi N$ and
 $\pi \pi$ interactions by using dispersion relations, while the Bonn
 potentials
 include both single-meson exchange ($\pi$, $\omega$, $\delta$) and the TPE
contributions ($\pi N\Delta$, $\pi\Delta\Delta$, $N\rho$ couplings) as
general meson-field description.
In Fig.1, an average $NN$ potential for $S$ wave is shown.
\begin{figure}[tb]
\epsfysize=2.50in
\centerline{\epsfbox{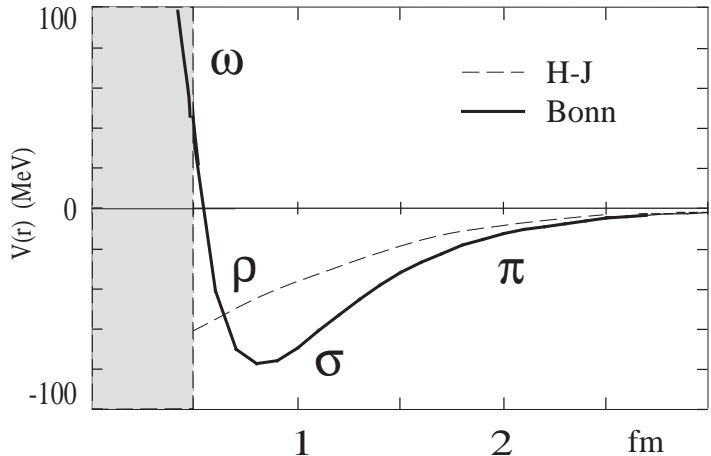}}
\caption{Average $S$ wave $NN$ potentials.}
\label{fig:hard-soft}
\end{figure}
The short-range repulsive terms have up to now been
treated phenomenologically
in all potential models. Since the middle 60's, the hard core is replaced
by a soft-core. These forms are demanded to fit the experimental
nucleon-nucleon phase shifts. A theoretical derivation of
form factor functions for the potential
near the origin would require further information on the subnucleonic structure
and even explicit modelling of the quark-gluon degrees of
freedom, which is beyond our current
understanding of quark physics.
Quantitatively, the contribution of OPE is about 95\% of the total
potential contribution for many nuclear properties,
as is the case for the deuteron; in a sense
the terminology ``realistic potential'' infers the realistic
treatment of OPE and TPE.

More specifically, in the Bonn potentials,
all the coupling constants are determined by
fitting $np$ scattering data and the deuteron properties.
It was claimed recently that the
potential models which fit better the $np$ data give poorer
agreement for the $pp$ data \cite{Nijmegen}. From deuteron properties
(i.e., the simplest $np$ system), we already know that generally
the $np$ components (i.e., mainly T=0 states) of the nucleon-nucleon
interaction are more attractive
than the $pp$ or $nn$ (i.e., mainly T=0 states) components.
Thus fitting only to $np$ scattering data will create a two-nucleon
potential which could overestimate the attractive components of
the $NN$ force with
respect to a unified $np$, $pp$ and $nn$ many-body system.
Recently, the Nijmegen group are attempting
to provide a potential model by
fitting both the $pp$ and $np$ scattering data. The work to
derive an effective interaction with this interaction
is still in progress.

\section{Few-Body and Many-Body Methods}

There are many available modern potentials which are based on the analysis of 
$N \pi$ scattering data within some given energy regimes (for
example, up to 350 MeV, the threshold of $\pi$ production ) and the deuteron
properties. These potentials are the
starting points for a microscopic study of the nuclear dynamics.
To this end, the
few-body system is an ideal testing-ground of the physics considered.
Nowadays the methods used  to obtain the nuclear few-body
solutions (i.e., in configuration-space or momentum-space)
provide very consistent and nearly exact solutions
for the bound-state problem.
Therefore, the results of the few-body problems will more or less
free one from approximations and concentrate on the study of
the nucleon dynamics.

Solving the Faddeev-Noyes equations for bound states
with a given nucleon-nucleon potential
should provide us with the correct binding energy for such systems.
Hence to study three-body bound systems, such as $^{3}$H and $^{3}$He,
are particularly important since they are the smallest ones beyond the
deuteron for one to quantitatively say something about their
properties and the first set of nuclei for one to investigate the
importance of hree body interactions.
This seemingly simple goal has been pursued by
many nuclear physicists for many years, and yet to date,
there is still no definitive answer
(for a recent review, see \cite{gibson}). For example, in
Table 1 we have presented the benchmark results, which explicitly do not
include three-body forces, of the predicted
$^{3}$H binding energies based upon various modern realistic potentials
and as functions of the number of three-body channels.
Except for the Bonn potential,
the calculated binding energies based on the
various $NN$ potential models are about 1 MeV underbound
(the experimental binding energy is 8.536 MeV for $^{3}$H).
The nuclear {\em rms} charge radii computed with
realistic $NN$ potentials are consistently larger than the experimental
values. This discrepancy between experiment and theory is unnerving.

One plausible explanation for the above problem is the presence of
a three body force.
Table 1 shows that a Hamiltonian of Reid soft-core two-body force plus
the Tucson-Melbourne two-pion exchange three-nucleon force (TM3N)
\cite{TM3N} can indeed provide additional binding for the trinucleon
system. However, these results are highly
sensitive to the $\pi$NN form factor
and its cutoff mass $\Lambda$ \cite{sasakawa}.
For instance, in a 34-channel calculation, the RSC+TM3N result for cutoff
mass $\Lambda$=5.8~m$_{\pi}$ is 8.86 (MeV), while the binding energies
are 7.46 (MeV) and 11.16 (MeV) for $\Lambda$=4.1 and 7.1 respectively.
Therefore, the current status is that while
the correct binding energy can be reproduced
with the adjustment of the three-nucleon force,
it is done so in a phenomenological manner.
On a more fundamental level,
three-nucleon interactions have recently been studied by the
chiral invariant effective lagrangians \cite{3bf}.
This may be an encouraging direction to fundamentally understand and
determine the effective nuclear three-body force.

\begin{table}[tb]
\caption{$^{3}$H binding energy (in MeV).}
\begin{center}
\label{tab:1}
\begin{tabular}{c|rrrrrr} \hline\hline
 potential& $\phantom{a}$ &  RSC & Paris & Nijmegen & Bonn-r
          & $\phantom{aa}$ \\ \hline
benchmark & $\phantom{a}$ & 7.02 & 7.31  & 7.49     & 8.23
          & $\phantom{aa}$  \\ \hline \hline
 channel  & $\phantom{a}$ & 5    & 9     & 18       & 34
          & $\phantom{aa}$   \\ \hline
   RSC    & $\phantom{a}$ & 7.02 &7.21   &7.23      & 7.35
          & $\phantom{aa}$  \\ \hline
RSC+TM3N  & $\phantom{a}$ & 7.55 &8.33   &8.93      &8.86
          & $\phantom{aa}$  \\  \hline \hline
\end{tabular}
\end{center}
\end{table}

It has also been argued that
the aforementioned discrepancy between the experimental
and calculated binding energy
is due to the missing of underlying baryon-meson degree of freedoms.
The $\Delta$-isobar and pion degrees of freedom should play some
role in the trinucleon binding \cite{Sauer}. After all, the
nuclear dynamics is just a low-energy limit of the baryon-meson dynamics.
Sauer et al \cite{sauer1} estimates that the presence of a single
$\Delta$-isobar provides about 0.3 (MeV) binding and that the three-nucleon
force with an intermediate $\Delta$-isobar gives 0.9 (MeV) additional binding,
whereas 0.6 (MeV) is lost due to the two-nucleon dispersive effect.
Finally, a single $\Delta$ excitation produces about 0.6 (MeV) binding energy,
which is not adequate to match the correct binding energy.
While such effects can contribute to the binding energies of the
respective three body systems, they
are not adequate to fully resolve the discrepencies.
Another factor which may muddy the water further
is relativistic corrections, namely whether or not nuclear
spectroscopy is insensitive to the off-shell
part of the $NN$ interaction.

The transition from a nuclear few-body system to a
nuclear many-body system requires new methodologies
to obtain a solution since the
traditional few-body techniques no longer apply due to obvious
computational difficulties. In particular, the
nucleon-nucleon interactions are normally not used directly in
the nuclear many-body problem. Instead, effective
interactions either ``derived'' from or motivated by
the fundamental interactions are normally 
used in the many-body context \cite{kirson}.
Hence such interactions must necessarily be intimately linked to the
Hilbert space of the calculation. Perhaps the most important
ingredient of the many-body systems is to introduce
the concept of the {\em mean field}.
It is often sufficiently accurate to obtain a valid solution
of the many-body problem.
Based on the {\em mean field} assumption,
the many body Schr\"odinger equation for the nucleus can be rewritten as
\begin{eqnarray}
& &  H  \psi_{\lambda}( 1,2,\cdots, A)
= \left( \sum_{i=1}^{A} t_{i}
    + \frac{1}{2} \sum_{i\neq j =1}^{A} v_{ij} \right)
 \psi_{\lambda}( 1,2,\cdots, A)   \\
& &\hspace{3cm} =(H_{0} + H' )
  \psi_{\lambda}( 1,2,\cdots, A)
    = E_{\lambda} \psi_{\lambda}( 1,2,\cdots, A),
\label{eq:h}
\end{eqnarray}
with
\begin{eqnarray}
& & H_{0}  \phi_{\mu}( 1,2,\cdots, A)
=\sum_{i=1}^{A} \left(t_{i}+u_{i} \right)
 \phi_{\mu}( 1,2,\cdots, A)
=E_{\mu}   \phi_{\mu}( 1,2,\cdots, A)
\label{eq:h0}  \\
& & H'= \left(\frac{1}{2} \sum_{i\neq j =1}^{A} v_{ij}
-\sum_{i=1}^{A}u_{i} \right),
\label{eq:res}
\end{eqnarray}
where $t_{i}$ is the kinetic energy,
and $u_{i}$ is the mean field that the i-th nucleon feels in the
many-body system, i.e., $u_{i}$ represents an average
potential contributed by all nucleons in a many-nucleon system.
$H'$ is the residual interaction.

Basically, there are two ways to deal with the
above many-body equations:
One is to focus on determining an optimal single-particle
mean field $H_{0}$, namely, Hartree-Fock method. The other is to
concentrate on figuring out a realistic two-body residual interaction
$H'$ in order to reproduce the spectroscopy
of the many-body system. In the latter case, the technique assumes the solution
to be a configuration mixture of Slater determinants. The
solution divides neatly into two parts: the calculation of
few  body terms (one- and two- for most interactions) in
a harmonic oscillator basis, followed by the many-body solution
in second quantized form  of eq.(\ref{eq:h0}).
This is the conventional Shell-Model approach.
Prior to performing the Shell-Model calculation, one needs
to derive the microscopic two-body interactions for a given model space.

Both the Hartree-Fock and the Shell-Model approaches
can also be used in conjunction with phenomenological
interactions. In the first approach, the use of the
Skyrme interaction leads to manageable calculations. It was shown
that this interaction is related to the fundamental
interaction \cite{Negele}. On the other hand, for the Shell-Model,
an approach based on a fit of the matrix elements in
second quantized form \cite{brown} attemps to
differentiate the quality (or lack thereof) of the fit to
experimental data as coming from either the many body aspects
of the calculations or from the quality of the two body
interaction. These approaches are fundamental in many
aspects.

\section{Effective Nuclear Interaction for Shell-Model Calculations}
In the 60's, Kuo and Brown presented a classic example to
pertubatively derive the microscopic effective interaction
(i.e., Shell-Model reaction matrix elements) \cite{KB}.
Later, a more exact and systematic way to derive the
effective two-body interaction for
a given model space was developed, namely the Folded-diagram method.
This method is reviewed in \cite{folder}. There is also
a brief review on the $NN$ effective interactions in \cite{brief}.

For example, using the folded-diagram method, one can determine the
63 two-body matrix elements in the $s$-$d$ shell, which together with
the three single-particle energies are the essential input in
the $s$-$d$ shell microscopic calculation.
Fig.2 gives a comparison of those 63 matrix elements from Hamada-Johnston,
Paris and Bonn-A potentials.  It shows quite vividly that the
different $NN$ potentials can provide very consistent 2-body matrix elements.
Only for some matrix elements, especially in the $T=0$ case,  are there
some discrapencies. However, these small discrapencies can produce
substancial differences in the binding enengy calculations
as we will see in the next section.
\begin{figure}[tb]
\epsfysize=2.50in
\centerline{\epsfbox{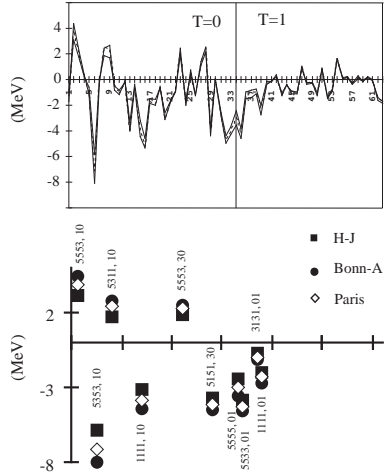}}
\caption{Comparison of microscopic two-body
matrix elements in the $s$-$d$ shell.}
\end{figure}

\section{Conventional Shell-Model Calculations}
Nuclei beyond the S-D shell require new techniques to handle the very
large model spaces involved. Using massive amount of CPU on the
fastest computers \cite{nakada} while keeping the traditional algorithms
is simply not
sufficient to model large nuclei. For this reason, there is renewed
interest in recent years to develop better, faster and more robust
Shell-Model algorithms. It deserves to be mentioned
in this direction are the pair
Shell-Model algorithm \cite{Chen} and the Monte-Carlo approaches
\cite{MC}. The former implements an exact Shell-Model solution for
even-even nuclei within a truncated Shell-Model space built from
arbitrary pair structures. It generalizes the FDU0 code \cite{fdu0}
code which was built from pairs envisioned in the Fermion Dynamical
Symmetry Model \cite{fdsm}. These codes can not
address directly the question of the worthiness of the fundamental
interaction since they seek a solution in a severely
truncated model space, which in itself would require an effective interaction
with uncertainties. The Monte-Carlo Shell-Model code on the
other hand seeks a solution of the many body problem via a Monte-Carlo
variational approach. It appears to work will in reproducing the
ground state properties of nuclei but suffers, at least up to now,
from a lack of convergence for realistic interactions \cite{MC}.

The Drexel University Shell-Model (DUSM) \cite{dusm} is another recently
completed code to perform Shell-Model calculations which embodies an entirely
new algorithm based on permutation group concepts. This is the code we used to
perform the calculations to be described in this report despites their
simplicity.
This code promisses great performance in cases where a full exact Shell-Model
 solutions are seeked for even-even, even-odd and odd-odd nuclei. It outperforms
standard Shell-Model codes \cite{OXBASH} by large (~4 or ~5) factors in CPU
usage with much reduced disk space and I/O requirements for $J$-$T$
(spin-isospin) coupled spaces.

\begin{figure}[tb]
\epsfysize=2.70in
\centerline{\epsfbox{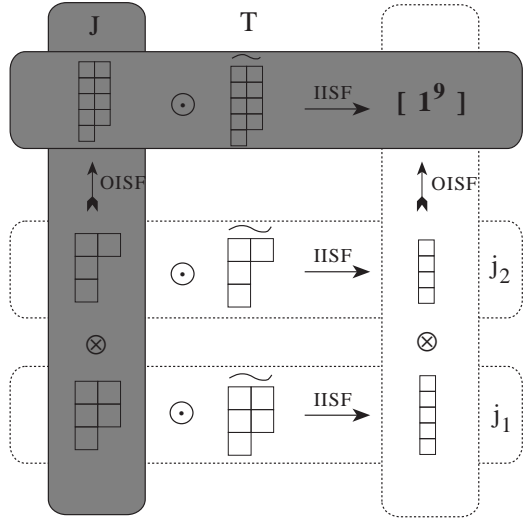}}
\caption{Two coupling schemes for multishell calculations.}
\label{fig:63}
\end{figure}

\subsection{Drexel University Shell Model Code (DUSM)}
The DUSM algorithm starts from the simple observation that Shell-Model
calculations are often ``multi-shell'' in only a single subspace. For
instance, a ``$J-T$''  calculation is multi-shell in the $J$ subspace but
single shell in the $T$ subspace ($\tau$=$\half$ for all nucleons). Or a
Shell-Model calculation for an electronic system is multi-shell in $L$ and
single-shell in $S$.  The wavefunctions in each subspace have arbitrary
permutational symmetry; these symmetry patterns are simply restricted to
be conjugate, so as to eventually couple to  total antisymmetric wavefunctions.
The DUSM coupling scheme is shown in the grayed area of Fig. 3.
The DUSM algorithm then proceeds to calculate wavefunctions, coupling
coefficents and matrix elements of elementary operators in each shell
for each subspace and then in the total space according to the
following steps:
\begin{enumerate}
\item Compute Coefficients of Fractioal Parentage (CFP) for each shell
in each subspace by diagonalysing the Casimir operator of $SU(2j+1)$ in
product basis.
\item Compute the matrix elements in all single states of the
elementary operators (all combinations of second quantized creation
and annihilation operators)
\item Compute the coupling coefficients  for the coupling among shells
(Outer Product Isoscalar Factors - OISF)
\item Compute the coupling coefficients  for the coupling between subspaces
(Inner Product Isoscalar Factors - IISF)
\item Compute the matrix elements of the Hamiltonian via a ``Sum over Path''
in permutation diagrams concepts
\item Diagonalyze the hamiltonian matrix
\item Loop over all total quantum number ($J$ and $T$ for instance) to
obtain the global solution
\end{enumerate}
The algorithm presents many advantages over traditional schemes: in particular
it avoids computing and storing total space CFPs. This implies much
reduced I/O during execution and much less disk storage. Single shell
calculations are known to be possible up to very large $J$ and number
of particle;
DUSM inherits this advantage. The DUSM algorithm uses group theory
concepts at all levels to obtain the coupling coefficients, diagonalising
the matrix represention of casimir operators of the appropriate groups at
all steps, implying very stable numerical schemes. This is an advantage
over using Racah formulas to derive the CFP \cite{Ji-Vallieres} in that
no explicit orthogonalization is ever needed.
The algorithm can furthermore be implemented
(with pointers) without any search since the approach
specifies fully the
range over all intermediate sums. The approach is fully coupled; it
provides full spectra and transitions among the states if required.

This code has recently been completed; we use it in this report to
compute some simple cases appropriate to understand the physics of the
$NN$ interaction. This does not illustrate by any means the
capabilities of DUSM.

\subsection{s-d Shell-Model Calculations}

We now describe calculations done in various model spaces via DUSM
using different effective interactions.

An important factor in the  comparison of
the experimental binding energies with results from large scale
Shell-Model calculations lies in the evaluation of the
Coulomb energy; this energy is  subtracted
from the total experimental binding energy.
The relation between total binding energy and the Coulomb term $E_{C}$
is as follow:

\begin{eqnarray}
& &  B_{total} = B_{core}  + B_{SM}  +  E_{C}^{SM},  \\
& &  E_{C}^{SM}= E_{C}^{total} - E_{C}^{core},
\label{eq:bind}
\end{eqnarray}
where $B_{core}$ is the binding energy for the core.
$B_{SM}$ is the Shell-Model binding energy related to the
core;  this term contains the binding energy of the cluster
valence-particle plus the binding between the cluster and core.
$B_{C}$ is the Coulomb energy.

The total binding energy and core binding energy are listed in
the 1993 atomic mass evaluation \cite{mass}.  Therefore,
according to eq.\ (\ref{eq:bind}), the binding energy calculated
from Shell-Model can be directly
compared with experimental values $B_{SM}$ when the Coulomb
term $E_{c}^{SM}$ is extracted from the experimental values.
In ref.\cite{WB}, the Coulomb energy is estimated with two assumptions:
first, for nuclei with a given proton number $Z$,
the Coulomb energies are independent of mass,
\begin{equation}
E_{C} (Z,A) = E_{C} (Z,A')~~.  \label{eq:rule1}
\end{equation}
Secondly, for nuclei with same mass $A$,
the Shell-Model binding energies measured with
respect to the analogue isospin states are the same
due to the isospin symmetry for the
interactions between identical nucleons (i.e., proton-proton force and
neutron-neutron force); consequently, $B_{SM}$ for mirror nuclei are the
same, while it differs from odd-odd nuclei due
to the excitation energy measured with respect to the
ground state,
\begin{equation}
E_{C}^{JT} (Z,A)  = E_{C}^{JT} (A-Z,A)~~.
\label{eq:rule2}
\end{equation}
For instance, when $^{16}O$ is considered as an inert core,
$B_{SM}(^{18}O)$=$B_{SM}(^{18}Ne)$
(the ground state is (JT)=01), while
for $^{18}F$, the excitation energy of the (JT)=01
 state with respect to its ground
state is 1.042 (MeV),
thus $B_{SM}(^{18}F)$ + 1.042 = $B_{SM}(^{18}O)$.
Notice that this way to estimate the Coulomb energies is not unique.
Ref.\cite{WB} claimed that the energy difference obtained via
different routes
is $\leq 150 keV$ under the above two assumption.

In more realistic formulation, the Coulomb energy should be closely
related to the proton distribution, which in turn
should be affected by the neutron distribution.
To be more specific, for example,
with more neutron outside the core, the attractive
interaction between proton and neutron tends to pull the protons towards
the surface (i.e., disperse the charge distribution),
thus reducing the Coulomb energy.
Therefore for nuclei with the same $Z$ but different neutron
number $N$, their $E_{C}$ should be  different. In fact, such
effect is expected to be crucial for neutron rich nuclei.

Alternatively, one can estimate the average Coulomb energy using an
empirical formula \cite{EC}
\begin{equation}
E_{C} = 0.717 \frac{Z^{2}}{A^{1/3}} (1-\frac{1.69}{A^{2/3}}) ~~(MeV)
\end{equation}
The second term in equation above is the Coulomb exchange term which takes into
account the effect of dispersion of nucleon due to Pauli exclusive effect.
We mention that the theoretical results from  the Fermi gas model
agree nicely with this empirical formula.  Recently, the finite-range
droplet model (FRDM) \cite{EC} was used to successfully
evaluate nuclear ground-state
masses and other nuclear structure properties\cite{EC},
including nuclei between the proton neutron drip
lines, showing that the error did not increase with distance from $\beta$
stability. Similar to the Strutinski's method,
FRDM describes nuclear binding energy
with two components, one is the so-called macroscopic energy, which
is,in some sense, taking care of nuclear bulk behavior, and the
second component models the microscopic energy correction
accounting for the nuclear shell
structures and two-body residual interaction (mainly the pairing effect).
The finite range effect of the nuclear force is
accounted for by the macroscopic part. Thus the Coulomb energy is further
modified \cite{EC}.
In table 2, we give the empirical Shell-Model binding energies of
$^{18}$O, $^{18}$F and $^{18}$Ne using different methods \cite{WB,EC}.
Note that there are negative values
of $E_{C}^{SM}$, which means that $E_{C}(^{18}O) < E_{C}(^{16}O)$
due to charge polarization.

\begin{table}[tb]
\caption{ The empirical Shell-Model binding energies of ground states.}
\begin{center}
\label{tab:2}
\begin{tabular}{cc|rrrrr}   \hline\hline
     &  (MeV)   &$\phantom{a}$&  $^{18}O$  & $^{18}F$  & $^{18}Ne$
&$\phantom{a}$ \\ \hline
 WB  &$B_{SM}$  &$\phantom{a}$&  -12.188   & -13.23    & -12.188
& $\phantom{a}$  \\
   &$E_{c}^{SM}$&$\phantom{a}$&  0         & 3.48      &  7.666
&  $\phantom{a}$   \\  \hline
 FRDM\footnote{$E_{c}(^{16}O)=11.185 (MeV)$}
    &$B_{SM}$   &$\phantom{a}$&  -11.9855   & -12.7966      & -11.2471
&$\phantom{a}$ \\
  &$E_{c}^{SM}$ &$\phantom{a}$&  -0.2025  & 3.0466  &  6.7251
& $\phantom{a}$  \\  \hline
\end{tabular}
\end{center}
\end{table}

In Table 3, we list the single particle energies for A=18 nuclei.
With the s.p. energies and 2-body matrix elements given from
sec.4, we now carry out the Shell-Model calculation for the
binding energies of ground states.
Table 4 presents the calculated binding energies from the microscopic
two-body matrix elements of Hamada-Johnston 
(i.e., Kuo-Brown (KB)
interactions which were derived perturbatively in \cite{KB}),
Bonn-A and Paris potentials.
Comparing with Table 2, one can see that KB reproduces
correct binding energies,
However, Bonn results overbind about 1.8 MeV for $T=1$ and 3.5 MeV
for $T=0$. Paris results are between KB and Bonn.
Notice that the 2-body matrix elements from Bonn and Paris potentials
are derived through a more precise method,
namely the folded-diagram approach.
These results of Shell-Model calculations are consistent with
the few-body calculations with three-body force, where Bonn
potential also gives rise to an overbinding.

\begin{table}[tb]
\caption{The s.p. energies of A=18 nuclei.}
\begin{center}
\label{tab:3}
\begin{tabular}{cc|rrrrr}   \hline\hline
& (MeV) &$\phantom{a}$&
  $d_{\frac{5}{2}}$ & $s_{\frac{1}{2}}$  & $d_{\frac{3}{2}}$
& $\phantom{a}$ \\ \hline
 $^{18}O$     &$E_{C}$=0  & $\phantom{a}$ &
  -4.144   & -3.273    & 0.941& $\phantom{a}$    \\
            &$E_{C}$=-0.103   & $\phantom{a}$ &
 -4.041       & -3.170      &  1.044 &$\phantom{a}$  \\ \hline
 $^{18}F$  &$E_{C}$=3.543  & $\phantom{a}$ &
 -4.144   & -3.649     & 0.856 &$\phantom{a}$    \\
         &$E_{C}$=3.179 &  $\phantom{a}$
 &  -3.780    & -3.285     &  1.220 $\phantom{a}$ &     \\  \hline
\end{tabular}
\end{center}
\end{table}

\begin{figure}[tb]
\epsfysize=3.00in
\centerline{\epsfbox{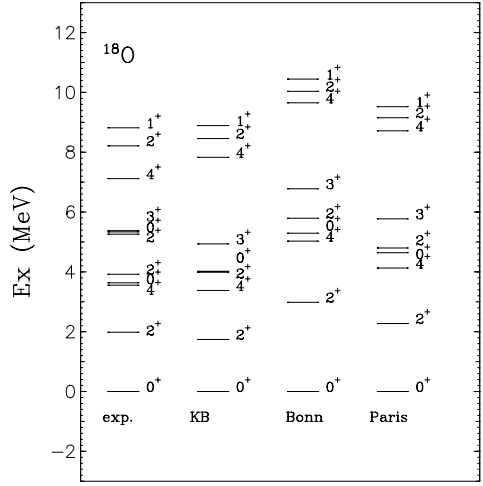}}
\caption{Spectra of $^{18}$O.}
\end{figure}
\begin{figure}[tb]
\epsfysize=3.00in
\centerline{\epsfbox{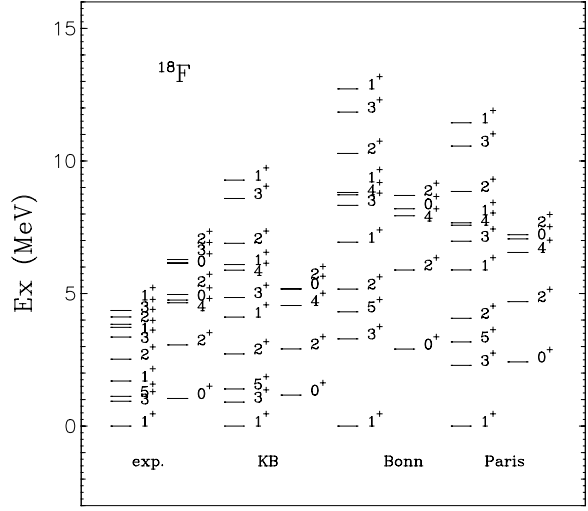}}
\caption{Spectra of $^{18}$F.}
\end{figure}

\begin{table}[tb]
\caption{ Comparison of binding  energies}
\begin{center}
\label{tab:4}
\begin{tabular}{c|rrrrrr}   \hline\hline
 Nuclei    &$\phantom{a}$&
 $E_{B_{exp}}$& KB (MeV) & Bonn (MeV) & Paris (MeV)& $\phantom{a}$ \\ \hline
 $^{18}O $ &$\phantom{a}$&
       kuo   & -12.20   & -13.96  &  -13.02 &  $\phantom{a}$  \\
           &$\phantom{a}$&
  $E_{C}$=0   & -12.19   & -13.95 &  -13.01 &$\phantom{a}$      \\
           & $\phantom{a}$&
 $E_{C}$=-0.103  & -11.98  &  -13.75   &  -13.02 & $\phantom{a}$    \\ \hline
 $^{18}F $ &$\phantom{a}$ &
 kuo & -13.36   & -16.86  &  -15.44 & $\phantom{a}$    \\
           &$\phantom{a}$ &
  $E_{C}$=3.543   & -13.58   & -17.09  & -15.68 & $\phantom{a}$     \\
           &$\phantom{a}$ &
  $E_{C}$=3.179   & -12.85   & -16.36 &  -14.95 & $\phantom{a}$    \\  \hline
 $^{18}Ne$ & $\phantom{a}$& &
            -12.20   &   -13.96   &   -13.02 & $\phantom{a}$   \\
         &$\phantom{a}$ &
  $E_{C}$=3.543   & -12.28   &  -14.07 & -13.12 & $\phantom{a}$ \\
         &$\phantom{a}$ &
  $E_{C}$=3.179   & -11.55   & -13.34  & -12.40 & $\phantom{a}$   \\ \hline
\end{tabular}
\end{center}
\end{table}

The above Shell-Model results for the binding energies are at best
confusing:
when one starts from a $NN$ soft-core potential, based on
effective meson theory with coupling constants being determined
by analyzing about a few thousands of $NN$ scattering data values and the
deuteron properties,
the binding energy results of the microscopic Shell-Model calculations
are overbinding. On the other hand, starting from a $NN$ hard-core potential,
which is given perturbatively (i.e., the early KB effective interactions),
the binding energy is reproduced very well.
It is not only that the the binding energy is somewhat better,
but the resulting spectra also demonstrate
that KB's presents the best agreement compared with results from other
modern potentials.
We illustrate this point in Fig.4 and Fig.5 where
 we show the theoretical and experimental spectra of
$^{18}$O and $^{18}$F.
It is worth mentioning that KB not only provides the best microscopic shell
model results for nuclei in $s$-$d$ shell; recently,
systematic Shell-Model calculations for $f$-$p$ nuclei have been
carried out by using the Kuo-Brown interaction for the $f$-$p$ shell
\cite{nakada} showing a similar trend.

  We may point out that a main difference between the early KB
effective interactions and the more recent ones
(for example, see \cite{brief})
is about the treatment of the folded diagrams. For the former the folded
diagrams were ignored, with the effective interaction given merely by
the bare-G and the second-order core polarization diagrams. These diagrams
are usually referred to as $G$ and $G_{3p1h}$ in the literature
\cite{sks83}. For the later, certain types of folded diagrams are
included to all orders using a $\hat Q$-box formulation \cite{sks83}.
What we have found in this work may indicate the need of a further
investigation of the folded diagrams. There may be some other physical
processes, which may counterbalance the effect of the folded diagrams
and which have not been investigated. When this is done, it may be possible
that the effective interaction will again be mainly given by
$G$ and $G_{3p1h}$ alone, making life simpler.

\section{Comments}
In summary, we have given a brief review of the
microscopic Shell-Model studies, and mentioned the recently developed
 Drexel University Shell Model (DUSM) code,
which implements a new Shell-Model algorithm.
We presented  simple Shell-Model calculations to
illustrate the progress made in this field
and discussed some significant problems which remain
in connection to the effective interactions used.

\begin{ack}
This work is supported
by the National Science Foundation.
\end{ack}

\end{document}